\documentstyle [12pt,titlepage]{article}
\newcommand{\be}{\begin{equation}}
\newcommand{\ee}{\end{equation}}
\newcommand{\lb}{\label}
\newcommand{\rf}[1]{(\ref{#1})}
\newcommand{\thalf}{\mbox{\small $\frac{1}{2}$\normalsize}}
\newcommand{\hs}{\mbox{$h_{s}$\ }}
\begin{document}
\title{Critical and off-critical properties of the $XXZ$ chain in
 external homogeneous and staggered magnetic fields}
\author{F. C. Alcaraz \thanks{e-mail: dfal@power.ufscar.br} and A. L.
Malvezzi\thanks{e-mail: palm@power.ufscar.br} \\
	Departamento de F\'\i sica \\ Universidade Federal de S\~ao Carlos \\
	 13565-905, S\~ao Carlos, SP, Brasil}
\date{PACS : 05.50.+q , 64.60.Fr , 75.10.Jm }
\maketitle
\voffset=-0.5cm
\textheight=19cm
\begin{abstract}
\vspace{0.2cm}
The phase diagram of the $XXZ$ chain under the influence of homogeneous
and staggered magnetic fields is calculated. The model has a rich phase
structure with three types of phases. A fully ferromagnetic phase, an
 antiferromagnetic phase and a massless phase with partial
ferromagnetic and antiferromagnetic order. This massless phase has
its critical fluctuations governed by a conformal field theory with central
charge $c=1$. When the $\sigma^{z}$-anisotropy $\Delta$ is zero the
model is exactly integrable through a Jordan-Wigner fermionization and our
results are analytic. For $\Delta\neq 0 $ our analysis are done numerically
using lattice sizes up to $M=20$. Our results shows that for
$-1 \geq \Delta \geq \frac{\sqrt{2}}{2}$ the staggered magnetic field
perturbation is relevant and for $1 \leq \Delta \leq \frac{\sqrt{2}}{2}$
it is irrelevant. In the region of relevant perturbations we show that
the massive continuum field theory associated to the model
has the same mass spectrum as the sine-Gordon model.
\end{abstract}
\section{Introduction}
\hspace{0.5cm} The anisotropic Heisenberg
model with spin $S=1/2$, or $XXZ$ chain, is one
of the most studied quantum spin system in Statistical Mechanics. Since
its exact integrability by Yang and Yang\cite{ref1} this model
is considered as the classical example of success of the Bethe ansatz. The
$XXZ$ chain give us the first example of a critical line with critical
exponents varying continuously with the anisotropy. More recently
\cite{ref2,ref3} with the advance of the conformal invariance ideas\cite{ref4},
 the whole operator content of this model was obtanied. The critical
fluctuations along this critical line is governed by a conformal field
theory with conformal anomaly $c=1$ and moreover their currents satisfy
 an $U(1)$ Kac-Moody algebra\cite{ref5}.

The effect of an uniform magnetic field in the phase diagram of the
$XXZ$ chain\cite{ref6} ( see Fig. 1 ) is to extend the critical phase
over a finite region delimited by the critical line of Pokrovsky-Talapov
 (P.T. line in Fig. 1)\cite{ref7},
 where the chain becomes fully ferromagnetic. Finite-size
studies of this extended massless phase\cite{ref8}
 showed that it is also described
by a $c=1$ theory with critical exponents now depending on the anisotropy
and magnetic field. In the antiferromagnetic region ( $\Delta < -1$
in Fig. 1 ), where the model is massive (noncritical), as the magnetic
field increases it destroys the massive phase entering again in the
massless phase.

Since the uniform magnetic field does not destroy the exact integrability
of the quantum chain, the eigenspectra in the whole extended massless
phase is exact solvable. This is not the case when we introduce other
perturbations, like a thermal perturbation ( which is related to a
dimerization in the model ) or an staggered magnetic field, where exact
integrability is lost. Some years ago Den Nijs\cite{ref9}, by
exploring the connection between the 8-vertex model and the
Gaussian model, conjectured
that a staggered magnetic field could be a relevant or irrelevant
perturbation depending upon the anisotropy strength. Consequently in
the region where this last perturbation is relevant we should expect an
apposite effect as that of the uniform magnetic field, since it will bring
the system into a massive phase. Motived by this we study in this paper
the phase diagram and critical properties of the $XXZ$ chain when both
magnetic fields are present. The $XXZ$ Hamiltonian with these fields is
given by
\be
\lb{H}
 H(\Delta,h,\hs)
 = -\frac{1}{2}\ \sum_{i=1}^{M}\  \left[ \ \sigma_{i}^{x}\sigma_{i+1}^{x} +
\sigma_{i}^{y}\sigma_{i+1}^{y} + \Delta\sigma_{i}^{z}\sigma_{i+1}^{z} +
2\left( \ h + (-1)^{i}h_{s} \ \right)\sigma_{i}^{z}\ \right] \ ;
\ee
where $\sigma_{i}^{x},\sigma_{i}^{y}$ and $\sigma_{i}^{z}$, $i=1,2,...,M$ are
Pauli matrices attached at the $M$ sites of the chain, $\Delta$ is the
$\sigma^{z}$-anisotropy parameter and $h,\ \hs$ are the uniform and
staggered fields, respectively. In this paper we assume periodic boundary
conditions in \rf{H}, and the lattice size $M$ as an even number.

Using all the machinery coming from finite-size scaling\cite{ref10} and
conformal invariance\cite{ref4}, we will calculate the critical lines,
exponents as well the masses appearing in the underlying massive
field theory around relevant perturbations.

The paper is organized as follows. In Section 2 we review the conformal
invariance relations relevant for our purposes and the operator content
in the absence of magnetic fields. In Section 3 we consider the cases
where $\hs \neq 0$. Initially we consider the case where $\Delta = 0$ or
 the $XY$ chain. This case is special since the
eigenspectra can be calculated analytically through a Jordan-Wigner
fermionization of the Hamiltonian. The general situation where
$\Delta \neq 0$ is studied numerically. In Section 4
we conclude with a general discussion of our results.
\section{Conformal invariance relations and the operator content when
$\hs = 0$ }
\hspace{0.5cm} Like most statistical mechanics
systems in $1+1$ dimensions we assume
that the Hamiltonian \rf{H} in its massless regime is conformally
invariant. Under this assumption for each operator\cite{ref4,ref11}
$O_{\alpha}$
 with dimension $x_{\alpha}$ and spin $s_{\alpha}$ in the operator
algebra of the infinite system, there exists an infinite tower of states
in the quantum Hamiltonian, in a periodic chain of $M$ sites, whose
energy and momentum as $M \rightarrow \infty$ are given by
\be
\label{towerE}
E_{j,j'}^{\alpha}(M) = E_{0}(M) +
\frac{2\pi v}{M}(x_{\alpha} + j + j' ) + o(M^{-1})
\ee
and
\be
\label{towerP}
P_{j,j'}^{\alpha}(M) = \frac{2\pi}{M}(s_{\alpha} + j - j' ) \ ;
\ee
where $j,j' = 0,1, ... $, $E_{0}(M)$ is the ground-state energy and $v$ is
the velocity of sound that can be determined by the dispersion relation
of the spectra or by the difference among energy levels belonging to the
same conformal tower.

The conformal anomaly $c$ can be obtained from the finite-size corrections
of the ground-state energy. For periodic chains, the ground-state energy
behaves asymptotically as\cite{ref12}
\be
\label{E0ccc}
\frac{E_{0}(M)}{M} = e_{\infty} - \frac{\pi c v}{6M^{2}} + o(M^{-2})
\ee
where $e_{\infty}$ is the ground-state energy per site in the bulk
limit.

The above relations has been applied successfully to a large number of
statistical mechanics systems. In the case of the $XXZ$ Hamiltonian with
 no external fields, $H(\Delta,h=0,\hs=0)$ and $-1 \leq \Delta \leq 1$ these
relations give us the whole operator content of the underlying field
theory\cite{ref2,ref3,ref5}. Along the critical line $-1 \leq \Delta \leq 1$
the central charge is $c=1$ and the anomalous dimensions appearing in the
model are given by integer numbers or by
\be
\lb{x(n,m)}
x_{n,m} = n^{2}x_{p} + \frac{m^{2}}{4x_{p}}
\ee
where $x_{p} = \frac{\pi - \cos^{-1}(-\Delta)}{2\pi}$ and $n$ and $m$ are
integers. The dimension $x_{0,1} = x_{p}$ is the same as that of the
polarization operator in the six vertex model. Moreover we can show\cite{ref5},
 by combining the integer dimensions with Eq. \rf{x(n,m)}, the model is
governed by a conformal field theory satisfying a larger algebra than
 the Virasoro conformal one, namely, a $U(1)$ Kac-Moody algebra. This
$U(1)$ symmetry comes from the commutation of the Hamiltonian \rf{H} with
 the $z$ component of the total spin\cite{ref13}
\be
\lb{Sz}
S_{z} = \frac{1}{2} \sum_{i}^{M} \ \sigma_{i}^{z} = n \ .
\ee
For a given sector of the Hilbert space where the total spin $S_{z} = n$
it will corresponds a set of primary operators $O_{n,m}$ with dimensions
$x_{n,m} \ (m=0,\pm 1, \pm 2, ...)$ and the number of descendants operators
(with dimensions $x_{n,m} + j + j' \ , \ j,j' \in \cal Z$) will be given by
the product of two $U(1)$ Kac-Moody characters\cite{ref5}. The operators
$O_{n,m}$ can be interpreted, in a Gaussian language\cite{ref14},
as the operators
with vorticity $n$ and spin wave excitation number $m$. The critical
 exponents are obtained from the dimensions \rf{x(n,m)}. For example the
exponents $\eta_{x}$ and $\eta_{z}$ governing the correlations
\be
\lb{correlx}
< \sigma^{x}(0)\sigma^{x}(r) > \ \sim \ r^{-\eta_{x}}
\ee
\be
\lb{correlz}
< \sigma^{z}(0)\sigma^{z}(r) > \ \sim \ r^{-\eta_{z}}
\ee
are related to the lowest eigenenergies in the sector with $S^{z} = 1$ and
$S^{z} = 0$, respectively. The exponents are given by
$\eta_{x}=2x_{0,1}=\frac{\pi - \cos^{-1}(-\Delta)}{\pi}$ and
$\eta_{z}=\eta_{x}^{-1}$ for $0\geq \Delta \geq -1$ and  $\eta_{z}=2$ for
$1\geq \Delta \geq 0$, since in this last region of anisotropies the lowest
 dimension above zero is one.

All the above results were obtained for $\hs=h=0$ and by exploring
the exact integrability  of the  model through the Bethe ansatz, which
numerically\cite{ref2} or analytically\cite{ref3} produced very precise
results. Such integrability is not destroyed by the introduction of the
uniform magnetic field and precise results can also be obtained in
this case. In the case where $\hs=0, \ h\neq 0$, as we see in Fig. 1,
the massless line extends forming a massless phase. This massless phase
is ordered ferromagnetically and for a given value of the
anisotropy $\Delta$ the magnetization changes from zero to its maximum
value where the system is fully ordered. the points where the magnetization
reaches its maximum  value form the Pokrovsky-Talapov transition
line\cite{ref7}, which separate the massive fully ferromagnetic phase from the
massless partially ordered ferromagnetic phase.

The physical mechanism producing the extended massless phase in Fig. 1
is due to the following. From Eqs. \rf{H} and \rf{Sz} the magnetic
field only decreases the eigenenergies in the sectors where $n>0$. In the
critical line $-1 \leq \Delta \leq 1$ and $h=0$ the ground-state belongs to the
sector $n=0$. Since we have no gap a small but finite positive magnetic field
will lead the ground-state into a sector with $n>0$, producing a partially
ordered ferromagnet phase. However the fluctuations around this ordered
phase are of the same nature as those appearing in the absence of the magnetic
field $(h=0)$. This argument imply that we should expect in the whole massless
phase a $c=1$ Gaussian field theory with the anomalous dimensions like
those given in Eq. \rf{x(n,m)}. In fact this fact was observed
analytically\cite{ref8} for $-1 \leq \Delta \leq 1$ and numerically\cite{ref15}
in the whole extended phase. In Refs. \cite{ref8}, although a closed
analytic form for the dimensions in the extended massless phase was not
derived,
the dimensions are shown to be of Gaussian type like in \rf{x(n,m)}, but
with $x_{p}$ now depending on $\Delta$ and $h$. The case where $\Delta=0$
is special since
only in this case the dimensions do not depend on the magnetic field h.

The finite-size analysis of the Hamiltonian \rf{H} with a nonzero uniform
magnetic field deserves some comments since for a fixed value of $h$ the
ground-state changes sector as the lattice size increase. In order to keep the
bulk limit physics $(M \rightarrow \infty)$ we should keep the magnetization
per particle fixed in the finite chains. This imply that for a given magnetic
field only a sequence of lattice sizes producing the fixed magnetization,
related to $h$, should be used.

The transition line of Pokrovsky-Talapov type in Fig. 1 is obtained by
calculating the lowest value of the magnetic field where the ground-state
is the ferromagnetic state ( sector $n=M/2$ ). This magnetic field is
$h=1-\Delta$ for arbitrary values of the lattice size $M$.

In the following sections we will calculate the effects where the staggered
 and the uniform magnetic field are included.
\section{Results for $\hs \neq 0$}
\hspace{0.5cm} We now analyze the situation
 where both magnetic fields $h,\hs$ are nonzero.
The staggered magnetic field for $h=0$  produces an antiferromagnetic ordered
ground-state. The ground-state belongs, as when $\hs=0$, to the sector
with $n=0$. In this section we are going to study analytically for
$\Delta = 0$, and numerically for arbitrary $\Delta$ the nature of the
resulting phase when \hs is introduced. Irrespective of the \hs value we
do expect that for sufficiently high values of $h$ $(h >> \hs)$ the same
mechanism discussed in last section which changes the ground-state sector
will also take place. Consequently we should have a phase transition surface
of Pokrovsky-Talapov type where the ground-state becomes the fully ordered
ferromagnetic state. The magnetic field $h(\Delta,\hs)$ at this surface,
as in the previous section, is calculated by imposing the ground-state as
the fully ordered ferromagnet state. This critical surface
\be
\lb{PT}
h = h_{c}^{PT} = \sqrt{h_{s}^{2} + 1} - \Delta \ .
\ee
is obtained by equating the lowest eigenenergy $E_{1} = -\frac{\Delta}{2}
(M-4) - h(M-2) -2\sqrt{h_{s}^{2} + 1}$, in  the sector with one spin down
with the ferromagnetic energy (zero spin down)
 $E_{0} = -(\frac{\Delta}{2} + h)M$.
For $h > h_{c}^{PT}$ the ground-state of $H(\Delta,h,\hs)$ is the ferromagnetic
 state. In order to study the cases where $h<h_{c}^{PT}$ we will consider
first the special case where $\Delta = 0$.
\subsection{$\Delta=0$}
\hspace{0.5cm} In this case
the Hamiltonian $H(\Delta,h,\hs)$ can be diagonalized through
a Jordan-Wigner transformation\cite{ref16}. This is done by introducing
fermionic operators
\be
\lb{JW}
c_{j} = \prod_{l=1}^{j} \ e^{i\pi \sigma^{+}_{l}\sigma^{-}_{l}} \
\sigma^{-}_{j} \ ,\hspace{0.8cm}
c_{j}^{\dagger} = \sigma^{+}_{j} \prod_{l=1}^{j} \
e^{-i\pi \sigma^{+}_{l}\sigma^{-}_{l} }
\ee
where $\sigma^{\pm}_{j} = \thalf (\sigma^{x}_{j} \pm i\sigma^{y}_{j})$ for
$j = 1,2,...,M$. It is easy to check that these operators satisfy the
fermionic algebra
\be
\lb{algebra}
\{ c_{j}^{\dagger},c_{j'} \} = \delta_{j,j'} \hspace{0.7cm}
\{ c_{j}^{\dagger},c_{j'}^{\dagger} \} =
\{ c_{j},c_{j'} \} = 0 \ .
\ee
In terms of these operators the conserved $U(1)$ charge is given by
\be
\lb{ferm.Sz}
S_{z} = \frac{1}{2} \sum_{i}^{M} \ \sigma_{i}^{z} =
\sum_{i=1}^{M} \ \sigma^{+}_{i}\sigma^{-}_{i} - \frac{M}{2} =
 \sum_{i=1}^{M} \ c^{\dagger}_{i}c_{i} - \frac{M}{2}
\ee
which imply that the Hilbert space associated to $H(0,h,\hs)$ can be
separated into block disjoint sectors labelled according to the fermionic
number
\be
\lb{Nf}
N_{F} = \sum_{i=1}^{M} \ c^{\dagger}_{i}c_{i} = \frac{2n+M}{2} \ .
\ee
For the sake of simplicity let us restrict to the case where $M$ is multiple
of 4, so that $N_{F}$ and $n$ have the same parity. For a given number of
fermions $N_{F}$ the Hamiltonian is given by
\begin{eqnarray}
\lb{Hfnd}
H(0,h,\hs) &=& \sum_{i=1}^{M-1} \ \left[ \ - \left( c^{\dagger}_{i}c_{i+1} +
c^{\dagger}_{i+1}c_{i}\right) + (-1)^{i}\hs\left( 2c^{\dagger}_{i}c_{i}
- 1 \right) \ \right] \nonumber \\
& &- 2nh + e^{i\pi N_{F}}\left(
 c_{M}^{\dagger}c_{1} + c_{1}^{\dagger}c_{M} \right) \ .
\end{eqnarray}
We see that the sectors with even and odd number of fermions are obtained
by antiperiodic and periodic boundary conditions respectively.

Since the Hamiltonian \rf{Hfnd} is quadratic in fermions operators we
can (see Appendix A of ref. \cite{ref16}), through a linear combination of
$c_{j},c^{\dagger}_{j}$ introduce new fermionic momentum operators
$\eta_{k},\eta^{\dagger}_{k} $ that brings the Hamiltonian into a diagonal
form
\be
\lb{Hfd}
H(0,h,\hs) = \sum_{k}\ \left( \ 2\theta_{k} \sqrt{h_{s}^{2} +
 \cos^{2}k} \ \right) \ \eta^{\dagger}_{k} \eta_{k} - 2nh
\ee
\be
\lb{thetak}
\theta_{k} =  +1 \hspace{.7cm} \mbox{for} \hspace{.7cm}
-\frac{\pi}{2} < k \leq \frac{\pi}{2}
\ \ \mbox{and} \ \ \theta_{k} = -1 \hspace{.7cm} \mbox{otherwize} \ ,
\ee
where the sum now runs over the momenta $-\pi \leq k \leq \pi$
and $\theta_{k}$ are signals\cite{ref17}.
Due to the boundary condition term in \rf{Hfnd} the possible values
of momentum $k$ will depend on the lattice size parity and are given by
\be
\lb{kNpar}
(2l + 1)\frac{\pi}{M} \ , \ \ l = -\frac{M}{2}, -\frac{M}{2}+1, . . . ,
 \frac{M}{2}-2, \frac{M}{2} -1
\ee
for $N_{F}$ even, and
\be
\lb{kNimpar}
2l\frac{\pi}{M} \ , \ \ l = -\frac{M}{2}, -\frac{M}{2}+1, . . . ,
 \frac{M}{2}-1, \frac{M}{2}
\ee
for $N_{F}$ odd.
The eigenenergies in a given sector with $N_{F}$ fermions are obtained
from the $\frac{M!}{N_{F}(M-N_{F})!}$ combinations of fermion energies.
The momentum of a given eigenstate is obtained by adding the fermions
momenta $P = \sum_{k} k$. For example the lowest eigenenergy in the sector
with $U(1)$ charge $n$ will be
\begin{eqnarray}
\lb{Eneven}
E_{n} &=& 2 \sum_{l=-\frac{M}{4}-\frac{n}{2}}^{\frac{M}{4}+\frac{n}{2}-1}
\ \theta_{k} \sqrt{h_{s}^{2} + \cos^{2}\left[ (2l+1)\frac{\pi}{M} \right]} -
2nh
\nonumber \\
&=& -2 \sum_{l=-\frac{M}{4}+\frac{n}{2}}^{\frac{M}{4}-\frac{n}{2}-1} \
\sqrt{h_{s}^{2} + \cos^{2}\left[ (2l+1)\frac{\pi}{M} \right]} - 2nh
\end{eqnarray}
for $n$ even and
\begin{eqnarray}
\lb{Enodd}
E_{n} &=& 2 \sum_{l=-\frac{M}{4}-\frac{n}{2}+\thalf}^{\frac{M}{4}+
\frac{n}{2}-\thalf}
\ \theta_{k} \sqrt{h_{s}^{2} + \cos^{2}\left(2l\frac{\pi}{M} \right)}
- 2nh
\nonumber \\
&=& -2 \sum_{l=-\frac{M}{4}+\frac{n}{2}+\thalf}^{\frac{M}{4}-
\frac{n}{2}-\thalf} \
\sqrt{h_{s}^{2} + \cos^{2}\left(2l\frac{\pi}{M} \right)}
 - 2nh ,
\end{eqnarray}
for $n$ odd. These energies corresponds to zero-momentum states since they
come from a symmetric distribution of fermion momenta.

{}From \rf{PT} we know that at $h = \sqrt{h_{s}^{2} + 1}$ the system undergoes
a
phase transition to the ordered ferromagnetic state. In order to detect other
phase transitions let us calculate the mass gap of $H(0,\hs,h)$ for small
values of $h$. In such cases the ground-state remains in the sector $n=0$
and the first excited state in the sector $n=1$. From Eqs. \rf{Eneven} and
\rf{Enodd} the mass gap is given by
\be
\lb{gap}
G = \lim_{M \rightarrow \infty} \ (E_{1} - E_{0}) = 2(h_{s} - h) \ .
\ee
Therefore, as long as $\hs > h$ the model is massive and ordered
 antiferromagnetically. At $h=\hs$ the model becomes massless and by the same
mechanism  we discussed in the last section we expect that as $h$ increases
 the model stay massless until the Pokrovsky-Talapov line is reached. This
intermediate phase has a ferromagnetic order since now the ground-state
change sectors as $h$ increases. In Fig. 2 we show the phase diagram
at $\Delta=0$.
\begin{center}
\bf{Intermediate phase at $\Delta=0$}
\end{center}

In order to analyze this phase using finite-size scaling we should keep the
magnetization per particle $\mu$ fixed. This is done by choosing a sequence
of lattice sizes $M_{1},M_{2},...$ and sectors $n_{1},n_{2},...$, such that
$\mu = \frac{n_{i}}{M_{i}} \ , \ i=1,2,...$ is fixed. For a given lattice
size $M_{i}$ and magnetization $\mu$ the magnetic field that brings the
ground-state in the sector where $n_{i}=\mu M_{i}$ should be in the range
$h_{min}(M_{i},\mu) \leq h \leq h_{max}(M_{i},\mu)$, where
$h_{min}(M_{i},\mu) = (E_{n_{i}}-E_{n_{i}-1})/2$ and $h_{max}(M_{i},\mu) =
(E_{n_{i}+1}-E_{n_{i}})/2$.
{}From Eqs. \rf{Eneven} and \rf{Enodd} we obtain in the bulk limit
\be
\lb{hmu}
h_{min}(\infty,\mu) = h_{max}(\infty,\mu) = h_{\mu} =
\sqrt{ h_{s}^{2} + \sin^{2}\alpha } \ ,
\ee
where $\mu = \frac{n_{i}}{M_{i}}$ and $\alpha = \mu \pi$.
Consequently if the applied field has the value
$h=h_{\mu}$ all the finite lattice sequence $M_{i} \ (i=1,2,...)$ will have
its ground-state in the sector with $U(1)$ charge $n_{i} = \mu M_{i} \
(i=1,2,...)$. It is simple to verify form Eqs. \rf{Eneven},\rf{Enodd} that
for arbitrary values of $0\leq \mu \leq 1$ the gap vanishes as $M \rightarrow
\infty$, and consequently this partially ordered phase is critical.
Assuming the critical fluctuations in this phase governed by a conformal
field theory, the relations \rf{towerE},\rf{towerP} and \rf{E0ccc} of
Sec. 2 enable us to calculate the anomalous dimensions and central
charge of this underline field theory.

For fixed value of $\mu$ the finite-size correction of the ground-state
 energy $E_{\mu M}^{GS}$ in the asymptotic regime $M \rightarrow \infty$
 can be obtained by using the Euler-Maclaurin formula\cite{ref18} in
Eq. \rf{Eneven}, which give us
\be
\lb{E0}
\frac{E_{\mu M}^{GS}}{M} = e_{\infty} \ -
\frac{\pi}{6M^{2}}\frac{\sin(2\alpha)}{h_{\mu}} + o\left( M^{-2}\right) \ ,
\ee
where
\be
\lb{einf}
e_{\infty}  =  -\frac{2}{\pi} \int_{0}^{\alpha}
\sqrt{ h_{s}^{2} + \cos^{2}t } \ dt \ - \ 2h_{\mu}\mu
\ee
is the energy per particle in the bulk limit $M \rightarrow \infty$. The
low lying excited states with nonzero momentum are calculated by changing
 the fermions momenta entering in the ground state. From
these energies we can derive the dispersion relation and the sound velocity
\be
\lb{zeta}
\zeta  =  \frac{\sin(2\alpha)}{h_{\mu}}
\ee
which changes continuously along the massless phase. By comparing Eqs.
\rf{E0} and \rf{zeta} with Eq. \rf{E0ccc} we see that the conformal anomaly
is $c=1$ for all values of $\mu$.

{}From Eq. \rf{towerE} the anomalous dimensions of the conformal operators
are calculated from the finite-size corrections of the mass gap amplitudes.
Such corrections, for a certain fermionic configuration, are calculated
straightforwardly by using the Euler-Maclaurin formula. For an arbitrary
 but rational magnetization $\mu = \frac{p}{q}$ (or magnetic field
$h=\sqrt{h_{s}^{2} + \sin^{2}\mu}$) the lattice sequence with size
$M_{i} = i q \ \ (i=1,2,...)$ will have its ground-state in the $U(1)$ sector
$n_{i} = \mu M_{i} = i p \ \ (i=1,2,...)$. The lowest eigenstate in neighboring
sectors where $n_{i}^{,} = n_{i} + \bar{n} \ \ (\bar{n} = \pm 1, \pm 2,...)$
has zero momenta and will give the mass gap
\be
\lb{massgap}
E_{n_{i}^{,}} - E_{n_{i}}^{GS} = \frac{2\pi}{M} \zeta \frac{\bar{n}^{2}}{4}
+ o\left( M^{-1}\right) \ .
\ee
The excited states in the sectors $n_{i}^{,} \ (i=1,2,...)$, obtained
by the addition (subtraction) of a momenta $j\frac{2\pi}{M}$
$\left( j'\frac{2\pi}{M} \right)$ to the fermions with positive (negative)
momenta will give us an eigenstate with momentum
\be
\lb{Pf}
P_{n_{i}^{,}}^{j,j'} = \frac{2\pi}{M} (j - j').
\ee
and mass gap
\be
\lb{Ef}
E_{n_{i}^{,}}^{j,j'} - E_{n_{i}^{,}}^{GS} = \frac{2\pi}{M}
\zeta \left( \ \frac{\bar{n}^{2}}{4} + j + j' \ \right)
+ o\left( M^{-1} \right) \ .
\ee
The relations \rf{towerE} and \rf{towerP}
 enable us to identify the eigenenergies
$E_{n_{i}^{,}}^{j,j'}$ as the conformal tower of the primary operator
with dimension $x_{\bar{n},0} = \frac{1}{4} \bar{n}^{2}$.

Other excited states in the sectors where $n_{i}^{,} = n_{i} + \bar{n}
\ (\bar{n} = \pm 1, \pm 2,...)$ are derived from the fermion configuration
producing $E_{n_{i}^{,}}^{j,j'}$. These eigenenergies are obtained by replacing
fermions with positive (negative) momenta by fermions with negative (positive)
momenta. The momenta of such states are
\be
\lb{Pf2}
P_{n_{i}^{,},\bar{m}}^{j,j'} = 2\pi \mu + \frac{2\pi}{M}(\bar{n}\bar{m} +
j - j')
\ee
and the mass gaps are
\be
\lb{Ef2}
E_{n_{i}^{,},\bar{m}}^{j,j'} - E_{n_{i}}^{GS} = \frac{2\pi}{M} \zeta
\left( \ \bar{n}^{2} x_{p} + \frac{\bar{m}^{2}}{4x_{p}} + j + j' \right)
+ o\left( M^{-1} \right)
\ee
where $x_{p}$ and $\bar{n},\bar{m},j,j' = 0, \pm 1, \pm 2,...$ . From
relations \rf{towerE} and \rf{towerP}  we identify Eq. \rf{Ef2} as the
conformal tower of the primary operator with dimension
$x_{\bar{n},\bar{m}} = \left( \frac{\bar{n}^{2}}{4} + \bar{m}^{2} \right)$.
Following the discussion of Sec. 2 ( see Eq. \rf{x(n,m)} ) we can interpret
$x_{\bar{n},\bar{m}}$ as the dimension of a Gaussian operator
 $O_{\bar{n},\bar{m}}$ with vorticity $\bar{n}$ and spin-wave number
$\bar{m}$. Furthermore, from the degeneracies of the low lying energies
in the conformal towers $\left\{ E_{n_{i}^{,},\bar{m}}^{j,j'} \right\}$ we
verify that their are given by the product of the
characters of $U(1)$ Kac-Moody algebra as discussed in Sec. 2 for the
case where $\hs=0$. Therefore the whole massless regime has the same critical
nature as in the case where $\hs=0$, being governed by a $c=1$ conformal
field theory with operators satisfying a $U(1)$ Kac-Moody algebra.

Comparing Eqs. \rf{Ef2},\rf{zeta} and \rf{x(n,m)} we see that while the
sound velocity depend on $h$ and \hs the dimensions $x_{\bar{n},\bar{m}}$
are constants. For example the correlations \rf{correlx} and \rf{correlz} are
given by
\be
\lb{ncorrelx}
< \sigma^{x}(0)\sigma^{x}(r) > \ \sim \ r^{-\thalf}
\ee
\be
\lb{ncorrelz}
< \sigma^{z}(0)\sigma^{z}(r) > - \mu^{2} \ \sim \ r^{-2}
\ee
where the magnetization per site $\mu$ is given by
\be
\lb{nmagne}
\mu = \frac{1}{\pi} \sin^{-1}\left( \sqrt{h^{2} - h_{s}^{2}} \right) \ .
\ee
As we shall see  this independence of exponents is a particular feature
of the case where $\Delta=0$. It is interesting to observe that contrary
to standard conformal towers appearing in critical systems,
Eq. \rf{Pf2} tell us that the eigenstates in the conformal tower has a
macroscopic momenta as long $\mu \neq 0$. This is due to the fact that
the fluctuations ruling the critical behavior are in top of a partially
ordered ferromagnet state. Another delicate point appearing here\cite{ref19}
 is that a continuum conformal theory can only be construct
for $h=\sqrt{h_{s}^{2} + \sin^{2}\mu}$, with $\mu$ having a rational
value, since the lattice sizes and sectors appearing in the finite-size
sequences are integer values $\mu = \frac{n_{i}}{M_{i}} \ \ (i=1,2,...)$.
For $h$ connected with  an irrational value of $\mu$ we can only obtain
a conformal theory by approximating $\mu$ by a close rational number.
\begin{center}
\bf{Massive phase at $\Delta=0$}
\end{center}

The antiferromagnetic phase for $h<\hs$ is massive with a mass gap given
by \rf{gap}. A continuum field theory describing the physics in this massive
phase can be obtained in the neighborhood of the perturbing parameter
$\delta = \hs - h
\put(5,2.5){$>$}\put(5,-3.3){$\sim$} \hspace{0.5cm} 0$.
Such field theory will be massive and the
masses can be estimated from the finite-size behavior of the eigenspectra.
The mass spectrum can be calculated by applying the scheme followed by
Sagdeev and Zamolodchikov\cite{ref20} in the study of the Ising model in
an external magnetic field. To do such calculations we should initially
find the finite-size corrections of the zero-momenta eigenenergies
$E_{k}(\delta,M), \ \ k=1,2,...$, at the conformal invariant
point $\delta = 0$. In the case of the Hamiltonian \rf{H} the results of
Ref. \cite{ref2}  tell us that such corrections, not only at $\Delta=0$
but for arbitrary values of $\Delta \ (-1\leq \Delta \leq 1)$, are governed
mainly by the irrelevant operator with dimension $\bar{x} = x_{0,2} =
1/x_{p}$ and the descendent of the identity operator with dimension 4. From
Ref. \cite{ref2} we have
\begin{eqnarray}
\lb{PertExp}
E_{k}(\delta=0,M) &=& e_{\infty}M + \frac{2\pi v}{M}\left( \ x_{k} -
\frac{c}{12} \ \right) + a_{1}\left(\frac{1}{M}\right)^{3} +
\\ \nonumber
& & a_{2}\left(\frac{1}{M}\right)^{\bar{x}-1} +
a_{3}\left(\frac{1}{M}\right)^{2\bar{x}-3} +
a_{4}\left(\frac{1}{M}\right)^{3\bar{x}-5} + ...
\end{eqnarray}
where $x_{k}$  is one of the dimensions \rf{x(n,m)} associated to
$E_{k}$ and $a_{1},a_{2},...$ are $M$-independent factors. According to
the scheme of Ref. \cite{ref20} if the perturbed operator which produces
the massive behavior has dimension $y$ we should calculate the eigenspectra
in the asymptotic regime $\delta \rightarrow 0\ , \ M \rightarrow \infty$,
with
\be
\lb{X}
X = \delta^{\frac{1}{2-y}} \ M
\ee
kept fixed. In this regime \rf{PertExp} is replaced by
\begin{eqnarray}
\lb{PertExp2}
E_{k}(\delta,M) &=& e_{\infty}M + \delta^{\frac{1}{2-y}}F_{k}(X) +
\delta^{\frac{1}{2-y}(\bar{x}-1)}G_{k}(X) +  \\ \nonumber
& & \delta^{\frac{1}{2-y}(2\bar{x}-3)}V_{k}(X) +
\delta^{\frac{1}{2-y}(3\bar{x}-5)}H_{k}(X) + ... \ .
\end{eqnarray}
The masses in the massive continuum field theory are obtained from the
large-$X$ behavior of the function \cite{ref20} $F_{k}(X)$, {\it i.e.},
\be
\lb{mass2}
m_{k} \sim F_{k}(X) - F_{0}(X)
\ee

The relations in Eqs. \rf{PertExp}-\rf{mass2} can be used for arbitrary
$\Delta$. In fact once we know $\bar{x} = 1/x_{p}$, the dimension
associated to the off-critical perturbation $y$ as well the generated
masses can be calculated from Eqs. \rf{X}-\rf{mass2}.

In the case $\Delta=0$ the perturbing parameter is $\delta = \hs - h
\put(5,2.5){$>$}\put(5,-3.3){$\sim$} \hspace{0.5cm} 0$.
By keeping $X=(\hs - h)^{\frac{1}{2-y}} \ M$ fixed, with $y$
unknown, we obtain from Eqs. \rf{Eneven} and \rf{Enodd}
\begin{eqnarray}
\lb{E1-E0}
E_{1} - E_{0} &=& 2(\hs - h) - \\ \nonumber
& & 2\sum_{l=1}^{X/2h_{s}^{\frac{1}{2-y}}}
\ \sqrt{ h_{s}^{2} + \sin^{2}\left[\frac{2\pi}{X}(2l)
h_{s}^{\frac{1}{2-y}}\right] } -
\sqrt{ h_{s}^{2} + \sin^{2}\left[ \frac{\pi}{X}(2l-1)
h_{s}^{\frac{1}{2-y}}\right] }
\end{eqnarray}
which gives for $\hs \rightarrow 0$
\be
\lb{gap3}
E_{1} - E_{0} \approx 2(\hs - h) - \frac{\Omega}{X}
h_{s}^{\frac{1}{2-y}}
\ee
where $\Omega$ is a constant. Consequently by comparing Eq. \rf{gap3}
with Eqs. \rf{PertExp2}-\rf{mass2} we obtain the dimension $y=1$ for
the perturbing operator and the mass $m_{1}=2$ associated to the
first gap.

The gap associated to the lowest eigenenergy in the sectors with $n$ even
is obtained from Eq. \rf{Eneven}
\be
\lb{Eneven-E0}
E_{n} - E_{0} = 4 \sum_{l=1}^{n/2} \ \sqrt{ h_{s}^{2} + \sin^{2}\left[
\frac{\pi}{X}(2l-1)h_{s}^{\frac{1}{2-y}}\right] } \ .
\ee
Since $n$ is finite, expanding for small values of \hs we get, by
comparing with Eq. \rf{PertExp2}, again the dimension $y=1$ for the perturbing
operator and
\be
\lb{Fn-F0}
F_{n}(X) - F_{0}(X) = 4 \sum_{l=1}^{n/2} \ \sqrt{1+
\frac{\pi}{X}(2l-1)^{2}}
\ee
which gives, for $X \rightarrow \infty$, the masses $m_{n} = n\ 2 = n\ m_{1}$.
The same result is also obtained in the case where $n$ is odd. Similar
calculations for zero-momentum (modulo $\pi$) excited states show us that
these states are associated with multiples of the mass $m_{1}$. However
our calculations shows the existence of two equal masses  in the system :
$m_{1} = m_{2} = 2$. This can be seen clearly by a two-fold
degeneracy of the first excited state with zero momentum (modulo $\pi$) in
the ground state sector $n=0$. The levels are related to the
threshold energy $2m_{1}$.

The dimension $y=1$ for the perturbing operator and the Gaussian dimensions
given in Eq. \rf{x(n,m)}, with $x_{p} = 1/4$, indicate that this operator
is the $O_{0,1}$ operator (see Sec. 2) with dimension $y=x_{0,1}=1$.
This imply that in the region where $x_{0,1} > 2$,
$(\Delta < \frac{\sqrt{2}}{2})$ this perturbation will be irrelevant and
the phase stays massless after the introduction of a small \hs .
\subsection{$\Delta \neq 0$}
\hspace{0.5cm} In this case the analytical
diagonalization done for $\Delta=0$ does not
work since the Hamiltonian through the Jordan-Wigner transformation
\rf{JW} will have a four-body fermionic interaction. Consequently our
analysis will be done numerically. The critical surfaces of
$H(\Delta,h,\hs)$, according to finite-size scaling theory (FSS)
\cite{ref10} can be obtained from the extrapolation $(M\rightarrow \infty)$
of the sequences $(\Delta^{(M)},h^{(M)},h_{s}^{(M)}) \ \ (M=2,4,...)$,
obtained by solving the equation
\be
\lb{FSS}
MG_{M}(\Delta,h,\hs) = (M-2)G_{M-2}(\Delta,h,\hs)
\ee
where $G_{M}(\Delta,h,\hs)$ is the gap of the Hamiltonian \rf{H} with
$M$ sites.

In order to simplify our analysis let us consider initially the plane
$h=0$. The critical surface is obtained by solving \rf{FSS} with
$\Delta$ or \hs fixed. In Table 1 and 2 we show some of the finite-size
sequences for lattice sizes up to $M=18$. In Table 1 \hs is fixed while
in Table 2 $\Delta$ is fixed. In Fig. 3 we show the curves in the
space $(\Delta,h)$, that solve Eq. \rf{FSS} for several lattice sizes.
We clearly see in the figure that as $(M\rightarrow \infty)$ the curves
for $\Delta < \Delta^{*} \approx 0.7$ tend toward $\hs = 0$, while for
$\Delta > \Delta^{*}$ the curves tend toward nonzero values of \hs. The
blow up region $\Delta \approx \Delta^{*}$ is also included in Fig. 3,
where the points we used to draw the continuum curves are also shown.
These results clearly indicate that for $\Delta < \Delta^{*} \approx 0.7$
the perturbation introduced by \hs is relevant producing a massive
antiferromagnetic phase, while for $1>\Delta>\Delta^{*}$ the model stays
massless until it reaches a critical staggered field $h_{s}^{*} =
h_{s}^{*}(\Delta)$ which depends on the value of $\Delta$. This is in
complete agreement with the predictions we did considering the $\Delta=0$
results, where we associated the operator with dimension
$x_{0,1} = 1/x_{p} = \frac{\pi}{2\left(\pi-\cos^{-1}(-\Delta)\right)}$ with the
staggered field perturbation. Therefore the perturbation is irrelevant
for $\Delta > \Delta^{*} = \frac{\sqrt{2}}{2} \approx 0.707$ and the phase
which appears for $\Delta > \Delta^{*}$ in Fig. 3 should be a massless
antiferromagnetic ordered phase, since $\hs \neq 0$ in this phase.
This is complete agreement with predictions did in an earlier work by
den Nijs\cite{ref9}.
The massless phase (denoted by ML in Fig. 3) is separated from the fully
ferromagnetic phase (denoted by FE in Fig. 3) by the Pokrovsky-Talapov
line (PT) $\hs = \sqrt{\Delta^2 -1}$, obtained from Eq. \rf{PT}.

We  confirmed numerically the massless nature of the ML phase. Using
relations \rf{towerE},\rf{towerP} and \rf{E0ccc} we verified that the
whole phase is governed by a $c=1$ Gaussian conformal field theory with
dimensions as in \rf{x(n,m)}, with $x_{p}$ varying continuously in the
whole phase. As a consequence of \rf{x(n,m)}, ratio of dimensions like
$x_{n,0}/x_{1,0} = n^{2}$ should be independent of $x_{p}$. In order to
illustrate this let us consider the ratio $G_{2}/G_{1}$ between the
gaps in the sectors $n=2$ and $n=1$. Since the gaps are related to
the dimensions $x_{2,0}$ and $x_{1,0}$, as $(M\rightarrow \infty)$ this
ratio should tend to the value $x_{2,0}/x_{1,0} = 4$. In Fig. 4
we show these ratios as a function of $\Delta$ for  $M=20$ and some
values of \hs. It is clear from this figure that the ratio changes to
the expected value of 4 at values of $\Delta$ which depend on the \hs value,
which is consistent with the results shown in Fig. 3.

Our results also show that the PT line of Fig. 3 has a different behavior than
the rest of the Pokrovsky-Talapov surface \rf{PT}, since only along this line
we
have a first order phase transition. When $h = 0$ by crossing the PT line
the ground-state changes from the ferromagnetic sector $n=M/2$ to the
sector $n=0$ discontinuously. As a consequence the magnetization per lattice
point changes discontinuously from 1 to zero, as we move from the FE
phase to the ML phase (see Fig. 3). Similarly the staggered magnetization
\be
\lb{stgmag}
\left\langle S_{z}^{*} \right\rangle = \frac{1}{2}
\left\langle \sum_{i=1}^{M} \
(-)^{i} \sigma_{i}^{z} \right\rangle
\ee
changes discontinuously from zero to a finite value, which depend on
$\Delta$. Also along the surface $PT$
the ground-state wave function which is nondegenerated for
$h \neq 0$ becomes $M$-degenerated when $h = 0$.
All the lowest states in the sectors with $n = -\frac{M}{2},-\frac{M}{2}+1,
...,\frac{M}{2}-1,\frac{M}{2}$ become degenerated. We can also show
that along this PT curve these degenerated ground-state wave function, for
a given sector $n$, is given exactly by
\be
\lb{wavefunc}
\Psi_{n}(\Delta,\hs,h=0) = \sum_{\{s\}} \,^{'} \ q^{\frac{1}{4}
\sum_{i=1}^{M} (-)^{i} s_{i}} \ |s_{1},s_{2},...,s_{M} \rangle
\ee
where
\be
\lb{com}
\Delta = \sqrt{h_{s}^{2} + 1} = \frac{q +
\frac{\textstyle 1}{\textstyle q}}{2} \ ,
\ee
$s_{i} = \pm 1 \ (i=1,2,...,M)$ are the eigenvalues in the $\sigma_{z}$ basis
and  the prime in the sum indicates that the configurations $\{ s \}$ are
constrained to $\sum_{i=1}^{M} s_{i} = n$.
\begin{center}
\bf{Massive phase at $\Delta \neq 0$}
\end{center}

Similarly as we did for $\Delta = 0$ let us now investigate the continuum
massive field  theory describing the fluctuations in the massive phase (AF) of
Fig. 3. The masses of this continuum theory are obtained from the
eigenspectrum of the Hamiltonian in the scaling regime given by Eq. \rf{X}
with $\delta = \hs$ and from our present analysis the dimension $y$ of
the perturbing operator is $y = x_{0,1} =
\frac{\pi}{2\left( \pi-\cos^{-1}(-\Delta) \right) }$. Using Eqs. \rf{PertExp},
\rf{PertExp2} and \rf{mass2} we can calculate the mass ratios of the continuum
theory. These are calculated from the asymptotic regime $M\rightarrow \infty$
 and $X\rightarrow \infty$ of the finite-size sequences
\be
\lb{ratios}
R^{(n)}_{k}(X,M) = \frac{F_{k}^{(n)}(X,M) - F_{0}^{(0)}(X,M)}
{F_{1}^{(0)}(X,M) - F_{0}^{(0)}(X,M)}
\ \longrightarrow \ \frac{m_{k}^{n}}{m_{0}^{1}}
\ee
The functions $F_{k}^{(n)}(X,M)$ are obtained by using in \rf{PertExp2}
 the finite-size sequence of the $k$ zero-momentum state $(k=0,1,2,...)$
in the $U(1)$ sector $n$. The associated mass is $m_{k}^{(n)}$ while the
lightest mass, obtained from the lowest eigenenergy in the sector with
$n=\pm 1$, is $m_{0}^{(1)}$. Our results indicate that for $\Delta \geq 0$
we have only two equal masses M$ = m_{0}^{(1)} =  m_{0}^{(-1)}$ and
a continuum starting at 2M. For $\Delta < 0$ other masses appears
depending on the value of $\Delta$. In Table 3 we present some of our
estimators for the mass ratios. We show for $M=20$ the ratios \rf{ratios}
for some values of $X$ and $\Delta$. Our results show the appearance for
$\Delta = - 1/2 $ of two masses $m_{1} = m_{0}^{(1)}$ and
 $m_{2} = m_{1}^{(0)}$ with a ratio $m_{2}/m_{1} = 1.61 \pm 0.03 $. For
$\Delta = -\frac{\sqrt{2}}{2} \ \left( \Delta = -\frac{\sqrt{3}}{2} \right)$
a third mass appears with the ratios $m_{3}/m_{1} = m_{0}^{(2)}/m_{0}^{(1)} =
2.0 \pm 0.2 \ (= 1.9 \pm 0.1) \ , m_{2}/m_{1} = 1.41 \pm 0.01\ (= 1.2 \pm
0.1)$.

It is well know that the six-vertex model (or the $XXZ$ chain) is transformed
into the eight-vertex model (or the $XYZ$ chain) by a thermal \linebreak
perturbation\cite{ref21}.
The spectrum of the transfer matriz of the eight-vertex model was calculated
\cite{ref22} explicitly and is the same as that of the sine-Gordon model
\cite{ref23}. From these results the thermal perturbation in the
$XXZ$ chain will produce the following sine-Gordon masses. For $\Delta \geq 0$
there will have a mass M twice degenerated plus a continuum starting at
2M. These are the soliton-antisoliton pair. For $\Delta<0$, beyond the
soliton-antisoliton pair bounded states will appear with masses depending
on $\Delta$,
\be
\lb{mass3}
m_{i+1} = 2\, m_{1} \,
\sin \left( \frac{\pi}{2} \frac{i}{\alpha}\right) \ ,
 \ i=1,2,...,[\alpha]
\ee
where
\be
\lb{comple}
\alpha = \frac{3\pi - 4\cos^{-1}(-\Delta)}{\pi}
\ee
and $[\alpha]$ is the integer part of $\alpha$.

Our results of Table 3 shows that also in the case where the perturbing field
is the staggered magnetic field the mass spectrum is the same as the
sine-Gordon model given by Eqs. \rf{mass3} and \rf{comple}.
\section{Conclusions}
\hspace{0.5cm} The results we obtained
in last section enable us to obtain a complete
schematic phase diagram of the Hamiltonian \rf{H}. For simplicity we only
draw the phase diagram for $\Delta \geq 0$ in Fig. 5. The surface
OGFHI separates the antiferromagnetic phase AF
 from the massless phase ML , while the surface
 AEDCB separates this last phase from the frozen ferromagnetic phase
FE. The order parameter characterizing such
phases is the magnetization
\be
\lb{mag}
\overline{M}_{h} =  \left\langle \ \frac{1}{2}\sum_{i=1}^{M}
\ \sigma_{i}^{z} \ \right\rangle
\ee
and the staggered magnetization
\be
\lb{stgmag2}
\overline{M}_{h_{s}} =  \left\langle \ \frac{1}{2}\sum_{i=1}^{M} \
(-)^{i} \sigma_{i}^{z} \ \right\rangle \ .
\ee
In phase FE
$\overline{M}_{h} \neq 0$ and $\overline{M}_{h_{s}} = 0$, in phase
AF  $\overline{M}_{h} = 0$
and \linebreak $\overline{M}_{h_{s}} \neq 0$, while for $h \neq 0$
 in phase ML we have
$\overline{M}_{h} \neq 0$ and $\overline{M}_{h_{s}} \neq 0$.
At $h=0$ phases AF and
ML  has only antiferromagnetic
order $\left( \overline{M}_{h} = 0, \ \overline{M}_{h_{s}} \neq 0 \right)$,
but these phases differs since contrary to phase
AF,
(or FE)
where the correlations have an exponential decay in the massless phase
ML the correlations has
a power-law decay.

All the phase transition surfaces of Fig. 5 are of second order except the
curve AE obtained when $h=0$. The surface AEDCB is the Pokrovsky-Talapov
surface given by Eq. \rf{PT}. In this whole surface the ground state is the
ordered Heisenberg state with all the spin up, except at the
line AE where the ground-state is $M$ degenerated with a wave function
given exactly by Eq. \rf{wavefunc}.

In the entire massless phase our numerical and analytical results show
that the critical exponents  change continuously with $h,\hs$ and $\Delta$,
except at $\Delta = 0$ where the exponents are constant. The critical
fluctuations in this phase is governed by a conformal field theory with
central charge $c=1$ and operators satisfying a $U(1)$ Kac-Moody algebra.

Our numerical and analytical analysis clearly indicate that the staggered
magnetic field perturbation is associated to the operator with dimension
$x_{0,1} = \frac{1}{4x_{p}} = \frac{\pi}{2(\pi - \cos^{-1}(-\Delta))}$.
This confirms an earlier prediction of den Nijs\cite{ref9} which states
that for $\Delta > \frac{\sqrt{2}}{2}$ this perturbation is irrelevant.

Coming from the massless phase
ML to the antiferromagetic
 phase AF, of Fig. 3 the
fluctuations will be ruled by a massive continuum field theory. Our analysis
of last section indicate that such massive field theory is the
sine-Gordon field theory with a mass spectrum that depends on the
value of $\Delta$ like in Eq. \rf{mass3}. This is the same mass spectrum
obtained by perturbing the $XXZ$ by a thermal field.
\begin{center}
\bf{Acknowledgments}
\end{center}
It is a pleasure to acknowledge profitable discussion with
R. K\"oberle, M. J. Martins and A. Lima-Santos.
This work was supported in part by Conselho Nacional de Desenvolvimento
Cient\' \i fico e Tecnol\'ogico - CNPq - Brasil.
\newpage
\begin{center}
\Large
Figure Captions
\normalsize
\end{center}
\vspace{0.7cm}

Figure 1 - Phase diagram of the Hamiltonian $H(\Delta,h,\hs)$ given in
Eq. \rf{H} for \hs = 0. The different phases and the Pokrovsky-Talapov
line are indicated.
\vspace{1cm}

Figure 2 - Phase diagram of the Hamiltonian $H(\Delta,h,\hs)$ given in
Eq. \rf{H} for $\Delta=0$. The  massless phase is separated from the
ferromagnetic phase by the  Pokrovsky-Talapov line $h = \sqrt{ h_{s}^{2}
 + 1}$ and from the antiferromagnetic phase by the line $h = \hs$.
\vspace{1cm}

Figure 3 -  Finite-size estimators of the phase diagram of the Hamiltonian
$H(\Delta,h,\hs)$ given in Eq. \rf{H} for $h=0$. The line are obtained by
solving Eq. \rf{FSS} for lattice size pairs $(M-2,M)$ $(M=10,12,...,18)$.
The antiferromagnetic, massless and ferromagnetic phases are indicated by
AF, ML and FE respectively. The Pokrovsky-Talapov line PT is also indicated.
The blow up region $0.57 \leq \Delta \leq 0.82$ is also
shown with the points used to draw the continuum curves.
\vspace{1cm}

Figure 4 - Ratios $G_{2}/G_{1}$ between the gaps $G_{2}$ in sector $n=2$
and $G_{1}$ in the sector $n=1$ of the Hamiltonian $H(\Delta,0,\hs)$
given in Eq. \rf{H}. The ratios are shown as a function of $\Delta$ for
lattice size $M=20$ and some values of \hs . In the region where the phase is
massless we expect the Gaussian relation $G_{2}/G_{1} \rightarrow
x_{2,0}/x_{1,0} = 4$ as $M \rightarrow \infty$.
\vspace{1cm}

Figure 5 - Schematic phase diagram of $H(\Delta,h,\hs)$ given in Eq. \rf{H}.
The phases indicated by FE, ML and AF are the fully ferromagnetic, massless
and antiferromagnetic phases, respectively. The Pokrovsky-Talapov surface
ABCDE separates the ferromagnetic and massless phases and the surface
FGOIH separates this last phase from the antiferromagnetic one. All the phase
transition surfaces are of second order excepting the line AE where the
transition is first order.
\newpage
\Large
\begin{center}
Table Captions
\end{center}
\normalsize
\vspace{2.0cm}

Table 1 - Sequences of estimators for the anisotropy $\Delta$ in the plane
$h=0$, obtained by solving Eq. \rf{FSS} with \hs kept fixed.
\vspace{1cm}

Table 2 - Sequences of estimators for the anisotropy $\Delta$ in the plane
$h=0$, obtained by solving Eq. \rf{FSS} with $\Delta$ kept fixed.
\vspace{1cm}

Table 3 - The mass-ratio estimators $R_{k}^{(n)}(X,M)$ defined in Eq.
\rf{ratios} for some values of the anisotropy $\Delta$ in the plane
$h=0$. The conjectured values of last line is given by Eq. \rf{mass3}.
\newpage
\topmargin=1.7cm
\textwidth=18.0cm
\textheight=22.0cm
\voffset=-3.2cm
\hoffset=-2.1cm
\Large
\begin{center}
Table 1
\normalsize
\vspace{1.cm}

\begin{tabular}{|c|c|c|c|c|c|c|c|}	\hline
$M-2,M$		&\hs = 0.01 &\hs = 0.03 &\hs = 0.05 &\hs = 0.1 &\hs = 0.2  &\hs = 0.5
&\hs = 1.0   \\ \hline
8,10		&-0.4540731 &-0.0159442 &0.3100008  &0.5969325 &0.7614057  &0.9864091
&1.3628443   \\
10,12		&-0.3507805 &0.2311949  &0.4642322  &0.6591962 &0.7855170  &0.9930197
&1.3634265   \\
12,14		&-0.2052818 &0.3724661  &0.5441739  &0.6931414 &0.7996157  &0.9972828
&1.3660457   \\
14,16		&-0.0457780 &0.4570385  &0.5922408  &0.7145697 &0.8088957  &1.0002445
&1.3650563   \\
16,18		&0.0891170  &0.5122369  &0.6242583  &0.7293916 &0.8154915  &1.0024008
&1.3657592   \\ \hline
\end{tabular}
\vspace{2cm}

\Large
Table 2
\normalsize
\vspace{1cm}

\begin{tabular}{|c|c|c|c|c|c|c|c|}	\hline
$M-2,M$		&$\Delta = 0.6$ &$\Delta = 0.65$ &$\Delta = 0.675$ &$\Delta = 0.7$
&$\Delta = 0.725$  &$\Delta = 0.75$ &$\Delta = 0.8$   \\ \hline
8,10		&0.1010376 &0.1213905  &0.1344105  &0.1499071 &0.1682486  &0.1890890
&0.2420444   \\
10,12		&0.0776255 &0.0958401  &0.1078859  &0.1225705 &0.1404855  &0.1621757
&0.2176779   \\
12,14		&0.0624300 &0.0789604  &0.0901779  &0.1041465 &0.1215976  &0.1432584
&0.2005065   \\
14,16		&0.0518302 &0.0669150  &0.0773753  &0.0906412 &0.1075649  &0.1290517
&0.1875885   \\
16,18		&0.0440532 &0.0578975  &0.0676763  &0.0802761 &0.0966522  &0.1178787
&0.1773956   \\ \hline
\end{tabular}
\newpage

\vspace{2cm}
\Large
Table 3
\normalsize
\vspace{1cm}

\begin{tabular}{|c|c|c|c|c|c|}	\hline
	& $\Delta = -\frac{1}{2}$ &
\multicolumn{2}{c|}{$\Delta = -\frac{\sqrt{2}}{2}$} &
\multicolumn{2}{c|}{$\Delta = -\frac{\sqrt{3}}{2}$} \\ \hline
   X	&$\frac{m_{2}}{m_{1}}$&$\frac{m_{2}}{m_{1}}$&
$\frac{m_{3}}{m_{1}}$&$\frac{m_{2}}{m_{1}}$&$\frac{m_{3}}{m_{1}}$ \\ \hline
   6	& 1.64115 & 1.41970 & 2.20667 & 1.25317 & 2.15167 \\ \hline
   10	& 1.61967 & 1.40502 & 2.06304 & 1.18785 & 2.01352 \\ \hline
   14	& 1.62223 & 1.40555 & 2.03037 & 1.15271 & 1.99746 \\ \hline \hline
Eq.\rf{mass3} & 1.61803 & 1.41421 & 2.00000 & 1.24698 & 1.94986 \\ \hline
\end{tabular}
\end{center}

\begin{thebibliography}{99}
%
%
\bibitem{ref1} C. N. Yang and C. P. Yang, Phys. Rev. {\bf 150},
321 (1966); {\bf 150}, 327 (1966).
%
\bibitem{ref2} F. C. Alcaraz, M. N. Barber and M. T. Batchelor,
Phys. Rev. Lett. {\bf 58}, 771 (1987); Ann. Phys. (N.Y.) {\bf 182}, 280 (1988).
%
\bibitem{ref3} C. J. Hamer, J. Phys. A {\bf 19}, 3335 (1986);
H. J. de Vega and F. Woynarovich, Nucl. Phys. {\bf B251}, 439
(1985); F. Woynarovich, Phys. Rev. Lett. {\bf 59}, 259 (1987).
%
\bibitem{ref4}  See, {\em e.g.}, J. L. Cardy in {\em Phase Transitions and
Critical Phenomena }, edited by C. Domb and J. L. Lebowitz
(Academic, New York, 1986) Vol. 11.
%
\bibitem{ref5}  F. C. Alcaraz, U. Grimm and V. Rittenberg,
Nucl. Phys. {\bf B316}, 735 (1989), J. Phys. A {\bf 21}, L117 (1988).
%
\bibitem{ref6} J. D. Jonhson and B. M. McCoy, Phys. Rev. {\bf A6},
1613 (1972); Takanishi, Progr. Theor. Phys. {\bf 50}, 1519 (1973);
 {\bf 51}, 1348 (1974); M. L\"uscher, Nucl. Phys. {\bf B117}, 475
 (1976); I. Affleck in {\em Fields Strings and Critical Phenomena},
 edited by E. Brezin and J. Zinn-Justin (North Holland, Amsterdam, 1990),
 Les Houches XLIX, p. 563.
%
\bibitem{ref7} V. L. Pokrovsky and A. L. Talapov,
Sov. Phys. JETP {\bf 51}, 134 (1980).
%
\bibitem{ref8} N. M. Bogoliubov, A. G. Izergin and V. E. Korepin,
Nucl. Phys. {\bf B275}, 687 (1986); F. Woynarovich, H-P Eckle and T.
 T. Truong, J. Phys. A {\bf 22}, 4027 (1989).
%
\bibitem{ref9} M. P. M. den Nijs, Phys. Rev. B {\bf 23}, 6111 (1981).
%
\bibitem{ref10} See, {\em e.g.}, M. N. Barber, in {\em Phase Transitions and
Critical Phenomena}, edited by C. Domb and J. L. Lebowitz (Academic,
New York, 1990), Vol. 8.
%
\bibitem{ref11} J. L. Cardy, Nucl. Phys. {\bf B270} [FS16], 186 (1966).
%
\bibitem{ref12} H. Bl\"ote, J. L. Cardy and M. Nightingale,
Phys. Rev. Lett. {\bf 56}, 742 (1986); I. Affleck,
ibid. {\bf 56}, 746 (1986).
%
\bibitem{ref13} The introduction of \hs and $h$ in the $XXZ$ Hamiltonian
does not destroy its commutation with the operator $S_{z}$.
%
\bibitem{ref14} L. Kadanoff and A. C. Brown, Ann. Phys. (N.Y.)
 {\bf 121}, 318 (1979).
%
\bibitem{ref15}  F. C. Alcaraz, M. Henkel and V. Rittenberg (unpublished).
%
\bibitem{ref16} E. Lieb, T. Schultz and D. Mattis, Ann.
Phys. (N.Y.) {\bf 16}, 407 (1961).
%
\bibitem{ref17} In the derivation of \rf{Hfd} from \rf{Hfnd}
the signals $\theta_{k}$ can be arbitrary. With our choice
all the excitations are combinations of particles.
%
\bibitem{ref18} See, {\em e.g.}, {\em Handbook of
Mathematical Functions}, edited by M. Abramowitz and I. A. Segun (Dover).
%
\bibitem{ref19} A similar problem where the continuum limit can only
be defined for certain values of the external parameters can be
find in \cite{ref8} and in F. C. Alcaraz and W. Wreszinski,
 J. Stat. Phys. {\bf 58}, 45 (1990).
%
\bibitem{ref20} I. R. Sagdeev and A. B. Zamolodchikov,
Mod. Phys. Lett. {\bf B3}, 1375 (1989).
%
\bibitem{ref21} See, {\em e.g.}, R. J. Baxter, {\em Exactly
Solved Models in Statistical Mechanics}, (Academic, New York, 1982).
%
\bibitem{ref22} J. D. Johnson, S. Krinsky and B. McCoy,
Phys. Rev. A {\bf 8}, 2526 (1973).
%
\bibitem{ref23} R. Dashen, B. Hasslacher and A. Neveu, Phys. Rev. D
{\bf 11}, 3424 (1975); A. B. Zamolodchikov and
A. B. Zamolodchikov, Ann. Phys. (N.Y.) {\bf 120}, 253 (1979).
%
\end{thebibliography}
\end{document}